\documentclass[preprint,superscriptaddress]{revtex4}
\usepackage{amssymb}

\usepackage{epsfig}

\begin{document}

\title{Constraint on nuclear symmetry energy from nuclear charge dependent neutron skin thickness of nuclei}

\author{Min Liu}
 \email{lium_816@hotmail.com}
 \affiliation{ College of Nuclear Science and Technology, Beijing Normal University, Beijing, 100875, P. R. China}
 \affiliation{College of Physics and Technology, Guangxi Normal University, Guilin, 541004, P. R. China}

\author{Zhuxia Li}%
 \email{lizwux@ciae.ac.cn}
 \affiliation{China Institute of Atomic Energy, Beijing, 102413, P. R. China}

\author{Ning Wang}%
 \affiliation{College of Physics and Technology, Guangxi Normal University, Guilin, 541004, P. R. China}

\author{Fengshou Zhang \footnote{Corresponding author : fszhang@bnu.edu.cn}}%
 \affiliation{ College of Nuclear Science and Technology, Beijing Normal University, Beijing, 100875, P. R. China}
 \affiliation{Beijing Radiation Center, Beijing, 100875, P. R. China}
\affiliation{Center of Theoretical Nuclear Physics, National
Laboratory of Heavy Ion Accelerator of Lanzhou, Lanzhou, 730000, P.
R. China}


\begin{abstract}
An alternative way to constrain the density dependence of the
symmetry energy with the slope of the neutron skin thickness of
nuclei which shows a linear relation to both the isospin asymmetry
and the nuclear charge with a form of $Z^{2/3}$ is proposed. The
linear dependence of the neutron skin thickness on the nuclear
charge and isospin asymmetry is systematically studied with the
data from antiprotonic atom measurement experimentally and the
extended Thomas-Fermi approach incorporated the Skyrme energy
density functional theoretically. An obviously linear relationship
between the slope parameter $L$ of the symmetry energy and the
slope of the neutron skin thickness on the isospin asymmetry can
be found by adopting 66 Skyrme interactions in the calculation.
Combining the available experimental data, the constraint of $16
\lesssim L \lesssim 66$ MeV on the slope parameter of the symmetry
energy is obtained. The Skyrme interactions satisfying the
constraint are selected.

\end{abstract}

\maketitle


The nuclear symmetry energy $S(\rho$) is the difference in energy
per nucleon between pure neutron matter and symmetric nuclear
matter, which is the key ingredient of the nuclear equation of
state(EoS) for asymmetric nuclear matter. It governs the important
properties of nuclei and neutron star. It also plays a significant
role in the nuclear reaction dynamics and the stability of the
phases within the neutron star and the interior cooling process
within it. It is well known that the density dependence of $S(\rho)$
for cold nuclear matter predicted from different models is extremely
variant. Acquiring more accurate knowledge of the density dependence
of the symmetry energy has become one of the main goals in nuclear
physics at present and in the near future and has stimulated many
theoretical and experimental
studies\cite{Tsang09,Chen05,Li08,Shet07,Fami06}. To characterize the
density dependence of the symmetry energy, $S(\rho)$ is expanded
near saturation density ($\rho_{0}$) as
\begin{equation}
S(\rho)=S(\rho_{0})+\frac{L}{3}(\frac{\rho-\rho_{0}}{\rho_{0}})
+\frac{K_{sym}}{18}(\frac{\rho-\rho_{0}}{\rho_{0}})^{2}+...,
\end{equation}
with the slope parameter
$L=3\rho_{0}\frac{dS(\rho)}{d\rho}|_{\rho_{0}}$ and the curvature
parameter
$K_{sym}$=9$\rho_{0}^{2}(\frac{d^{2}S(\rho)}{d\rho^{2}})|_{\rho_{0}}$.

The calibration of neutron skin thickness of nuclei defined by
      $\Delta R_{np} = \left<r^{2}_{n}\right>^{\frac{1}{2}} - \left<r^{2}_{p}\right>^{\frac{1}{2}}$
has attracted a lot of attention in recent years because of the
sensitivity of  $\Delta R_{np}$ to the density dependence of the
symmetry energy. The calculations in either non-relativistic or
relativistic mean-field models show a well-defined linear
correlation between the $\Delta R_{np}$ of heavy nuclei and the
slope parameter $L$ of the symmetry energy at the saturation density
\cite{Furn02,Avan07,Baldo04}. Thus, $\Delta R_{np}$ of nuclei can be
used as a powerful observable to constrain the density dependence of
the symmetry energy at $\rho_{0}$ and lower densities. The
difficulty in the calibration of neutron skin thickness of nuclei
stems from the difficulty in the measurement of neutron
distribution. The main methods to measure the neutron distribution
or neutron skin thickness include hadron
scattering\cite{Rayn79,Hoff81,Kell91,Stra94}, $\pi^{-}$ elastic
scattering\cite{Taka95}, antiprotonic
atoms\cite{Jast04,Trzc01,Lubi98,Lubi94}, excitation of giant
dipole\cite{Kras91,Satc87,Kras04} and spin-dipole
resonances\cite{Gaar81,Ange80} on inelastic alpha scattering.
Unfortunately, the obtained values of $\Delta R_{np}$ from different
experimental method depend on the used analysis model and sometimes
are not totally consistent with each other. It is also hard to judge
the model dependence of the systematic error in different
experimental method. The parity-violating electron
scattering\cite{Horo01,Horow01} will be a hopeful option to measure
the neutron distribution with unprecedented precise of 1\% in a
model independent way. However, it is still not available up to now.
In this case, it will be difficult to accurately and consistently
constrain the symmetry energy by directly using the data of neutron
skin thickness. We notice that the $\Delta R_{np}$ of 26 stable
nuclei all over the periodic table (from $^{40}$Ca to $^{238}$U)
have been accumulated from antiprotonic atom measurement. In
Ref.\cite{Jast04}, the dependence of $\Delta R_{np}$ on the isospin
asymmetry $\delta$ = (N $-$ Z)/A for these 26 nuclei was extracted
from experimental data of antiprotonic atom measurement, which reads
\begin{equation}
\Delta R _{np} =(-0.03 \pm 0.02)+(0.90 \pm 0.15) \delta.
\end{equation}
Recently, Warda et al. represented this relationship in droplet
model with surface width dependence\cite{Cent09}. However it is
already known that the antiprotons are only sensitive to the tail
of the neutron distribution. An assumed shape for the neutron
density is needed to extract the rms radius. Therefore the
uncertainty in the value of $\Delta R_{np}$ is unavoidable in this
approach.

In this letter, we suggest an alternative way to constrain the
density dependence of the symmetry energy by the slope of a nuclear
charge dependent linear relation of $\Delta R _{np}$ \textit{vs}
$\delta$ based on the 26 experimental data from the antiprotonic
atoms obtained up to now. In this way, only the tendency of $\Delta
R _{np}$ changing with the isospin asymmetry $\delta$ is required to
get the information of the symmetry energy. The systematic
uncertainty due to the experimental method itself is therefore
expected to be largely reduced. The more data of $\Delta R _{np}$
for different nuclei with the same experimental method, the more
accurate constraint on the density dependence of the symmetry energy
can be obtained. One should take into account the nuclear charge
dependent linear relation of $\Delta R _{np}$ \textit{vs} $\delta$.
Because the neutron skin thickness is the difference of the neutron
and proton rms radii, it should depend not only on the symmetry
energy but also on the Coulomb interaction which is far from clear.

Let us first study the systematic behavior of the nuclear charge
dependence of $\Delta R _{np}$. The approach of semi-classical
extended Thomas-Fermi approximation\cite{Vaut72,Brack85} up to the
second order (ETF2) incorporated a potential energy density
functional including the standard Skyrme energy density and Coulomb
energy density with the Coulomb exchange term is applied in the
calculations, in which the nuclear surface diffuseness is
self-consistently taken into account. We calculate the proton and
neutron density distributions of nuclei (See \cite{Lium06} for
details) with a spherical symmetric Fermi functions by means of
restricted density variational
method\cite{Bart85,Brack85,Cen90,Bart02}. With the density
distributions determined in this way, the ground state properties
such as the energy and the nuclear charge radii of series of nuclei
have been calculated. The corresponding experimental data can be
reasonably well reproduced\cite{Liumin06}. Based on the calculated
rms radii of proton and neutron, we get the neutron skin thickness $
\Delta R_{np}$. The effective Skyrme interaction SLy4 is adopted in
this work since SLy4 is very successful in describing the bulk
properties and surface properties of nuclei\cite{Ange04}.
 \begin{figure}
 \includegraphics[angle=-0,width=0.7\textwidth]{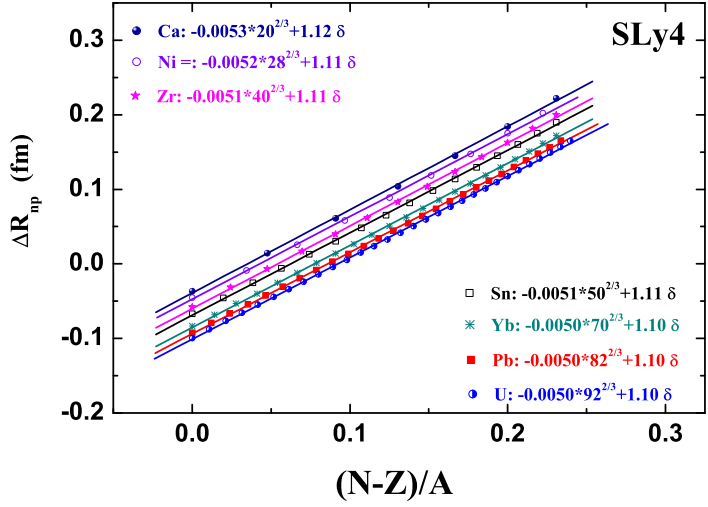}
        \caption{Linear correlation between the neutron skin thickness and the isospin asymmetry
         for Ca, Ni, Zr, Sn, Yb, Pb and U elements. Scatter symbols denote the calculation
         results. Solid lines denote the linear fitting results with a form
          $ C_{Z}  Z^{2/3} + C_{\delta} \delta$.}
 \end{figure}

  \begin{figure}
 \includegraphics[angle=-0,width=0.7\textwidth]{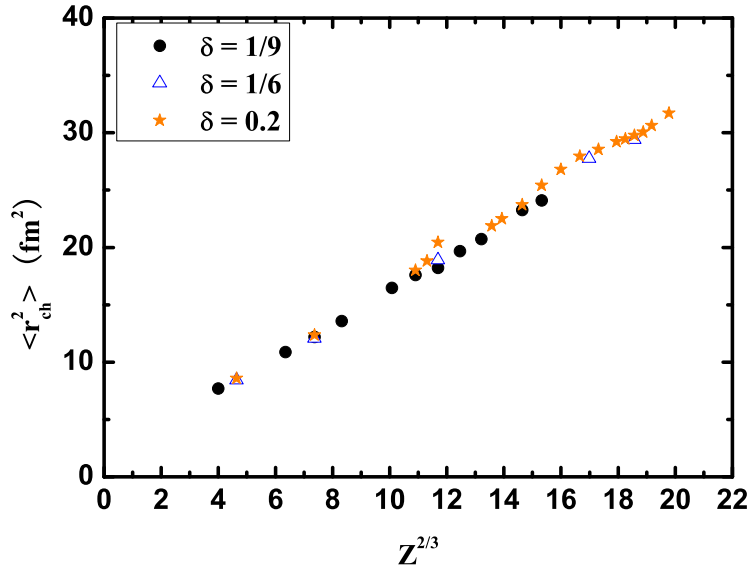}
        \caption{ Measured charge mean square radii $\left<r^{2}_{ch}\right>$  \cite{Ange04}
        as a function of $Z^{2/3}$ for selected nuclei with the same $\delta$.}
 \end{figure}

With this approach we calculate the $\Delta R_{np}$ for the
isotopes with charge number in the range of 20$\le$ Z $\le$92,
only the even-even nuclei are taken into account. Fig.1 shows the
correlations between $\Delta R_{np}$ and $\delta$ for Ca, Ni, Zr,
Sn, Yb, Pb and U isotopes. It is seen from the figure that the
$\Delta R_{np}$ and the isospin asymmetry $\delta$ of nuclei are
linearly correlated for Ca, Ni, Zr, Sn, Yb, Pb and U isotopes,
respectively. The fitting lines are nearly parallel with each
other and the slopes of $\Delta R_{np}$ as a function of $\delta$
for all selected elements is about 1.10 fm per unit $\delta$
within the range of 0 $\le \delta \le$ 0.24. The intercepts of the
lines depends on the nuclear charge and roughly has a form of
$\propto Z^{2/3}$. We have varied the form of the charge number
dependence from $Z^{1/3}$ to $Z^{4/3}$ and find that only the form
of $\propto Z^{2/3}$ can describe the charge dependence of $\Delta
R_{np}$. To understand the $\propto Z^{2/3}$ dependence of $\Delta
R_{np}$, we investigate the relation between the squared charge
mean radii and the charge number Z for the nuclei with same
isospin asymmetry. We find that the squared charge mean radii for
the nuclei with same isospin asymmetry relate to $Z^{2/3}$
linearly. Fig.2 shows a monotonic increasing tendency of the
experimental data of $\left<r^{2}_{ch}\right>$ \cite{Ange04} with
$Z^{2/3}$ for nuclei with $\delta  = \frac{1}{9}$, $\frac{1}{6}$,
and $0.2$, respectively. Obviously, increasing the charge number
for the nuclei with same isospin asymmetry enlarges  the
$\left<r^{2}_{ch}\right>$ and thus reduce the neutron skin
thickness.

 \begin{figure}
 \includegraphics[angle=-0,width=1.0\textwidth]{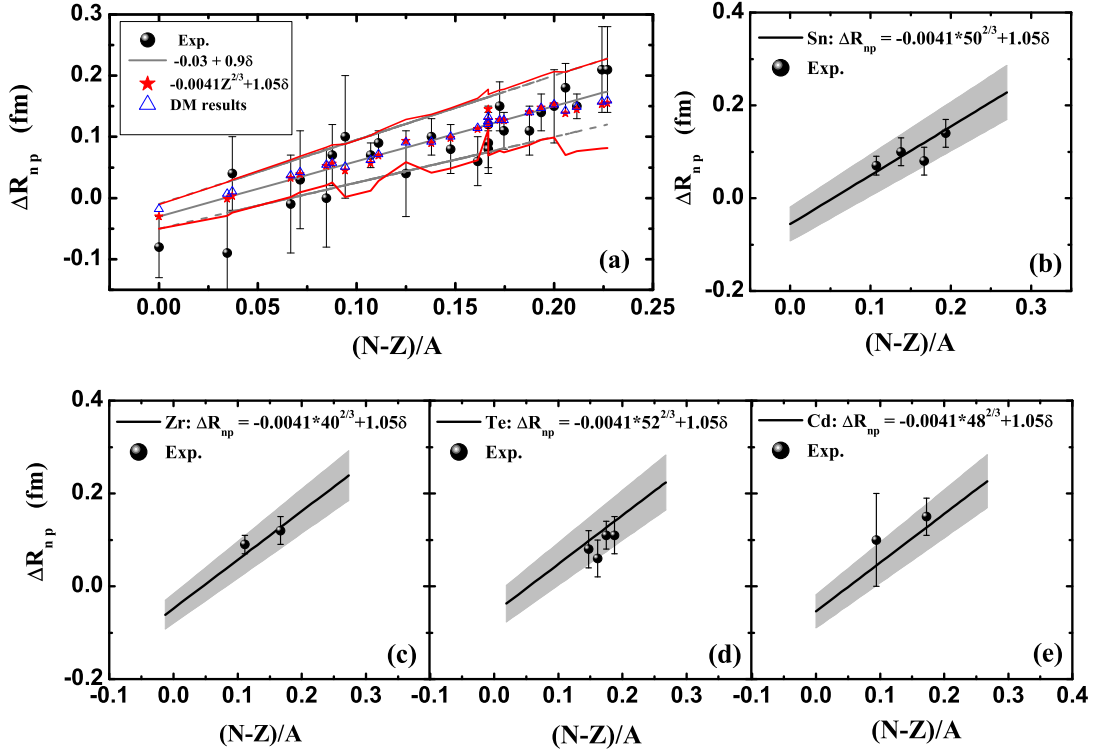}
        \caption{(Color online) (a)Comparison of the results with $(-0.0041 \pm 0.0027) Z^{2/3} + (1.05 \pm 0.08) \delta$
        to the experimental neutron skins from antiprotonic measurements\cite{Jast04}.
        Results of DM are also shown. The dashed gray lines denote the upper and lower limits
        of Eq.(2) and the solid red lines denote those of Eq.(4).
        (b) to (e) Comparison of the results with the fitting linear
        formula to the experimental data from antiprotonic measurements for Sn, Zr, Te and Cd isotope chains. }
\end{figure}

According to the above investigation, we propose an empirical
expression for the description of the neutron skin thickness of a
nucleus as
    \begin{equation}
       \Delta R_{np} =C_{z} Z^{2/3} + C_{\delta} \delta.
    \end{equation}
We have  performed a $\chi^{2}$ analysis in order to obtain the
optimized $C_{z}$ and $C_{\delta}$ from the experimental data of
$\Delta R_{np}$ from antiprotonic atom measurement. But the
accumulated data of $\Delta R_{np}$ are not enough to get a precise
constraint on the parameters. Thus we obtain the parameters $ C_{Z}$
and $C_{\delta}$ by fitting the expression (2) obtained from
experimental data of antiprotonic atom measurement. Firstly we use
the data $\Delta R _{np}$ for $^{40}$Ca to determine the range of
$C_{Z} = (-0.0041 \pm 0.0027)$fm because for $^{40}$Ca the $\delta$
is equal to zero. Then we adopt $ C_{Z} = -0.0041$ fm to fit $\Delta
R_{np}$ for all 26 experimental data available up to now, in which
the weight from error bar are taken into account to obtain the
optimized $C_{\delta}$. The fitting $C_{\delta}$ of 1.05 fm is
obtained. The range of $C_{\delta} = (1.05 \pm 0.08)$ fm is
determined through two limits in the expression (2) extracted from
experimental data. We finally obtain the neutron skin thickness as a
function of both the nuclear charge and the isospin asymmetry of
nuclei, extracted from the available data of $\Delta R_{np}$ given
by antiprotonic atom measurement as
\begin{equation}
 \Delta R _{np} =
(-0.0041 \pm 0.0027)Z^{2/3}+(1.05 \pm 0.08) \delta.
\end{equation}
The uncertainty of $C_{\delta}$ determined in this approach is
considerably reduced from $0.15$ fm  in Eq.(2) to $0.08$ fm when the
charge dependence of neutron skin thickness is taken into account.
The comparison between the $\Delta R _{np}$ obtained with this
expression and  experimental data is given in Fig.3(a). The results
of droplet model (DM)\cite{Cent09} are also shown in Fig.3(a) for
comparison. One can see that Eq.(4) can reproduce most of the
experimental data well and also is in good agreement with the DM
calculations. In the sub-Figs (b),(c),(d),(e) of Fig.3, we pick out
Sn, Zr, Te, Cd isotope chains from the available 26 experimental
neutron skin thickness data and compare with the results of Eq.(4).
One can see that the experimental data are reproduced reasonably
well.

 \begin{figure}
\includegraphics[angle=-0,width=1.0\textwidth]{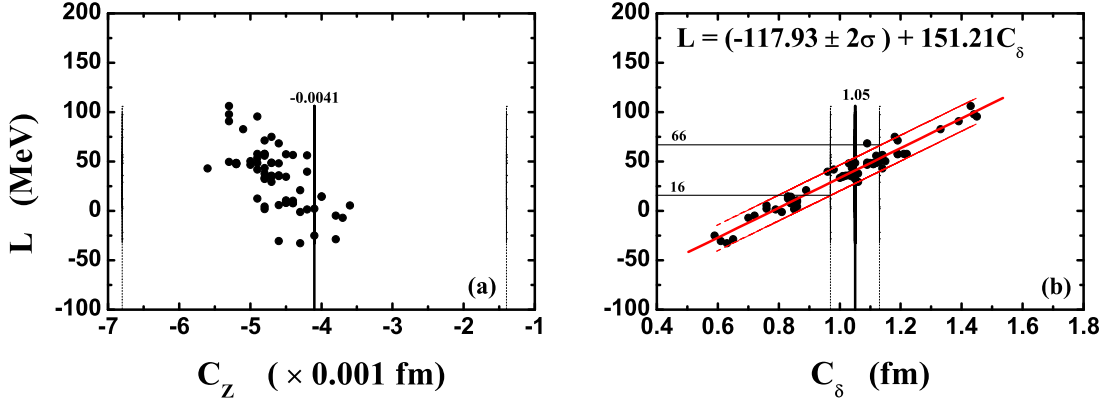}
        \caption{(Color online) The correlation between $L$ and the parameters $C_{Z}$ and $C_{\delta}$, respectively.
         The full circles are the calculated $ C_{Z}$ and $C_{\delta}$ for each Skyrme interaction. The solid black vertical lines
         denote the extracted value of $ C_{Z}$ and $C_{\delta}$ from experiment. The dashed black vertical lines denote
         the extracted upper and lower limits of them.}
\end{figure}

Now let us come to explore the density dependence of the nuclear
symmetry energy from the neutron skin thickness of nuclei $\Delta
R_{np}$ correlated with both charge and $\delta$ of nuclei. We
perform a systematic calculations within the ETF2 approach by using
66 sets of Skyrme interactions like SLy series, SkT~series,
$v$~series, Skz0$\sim$4, MSk1$\sim$6, SkSC1$\sim$4, SkM, SkM$^{*}$,
SkM1, SkMP, SII $\sim$ SVII, SIII$^{*}$, SGI, SGII, Zs, Es, Gs, Rs,
FitB, BSk1, RATP, SKRA. The slope parameter $L$ of the bulk symmetry
energy covers a widely range from nearly $-30$ MeV to $100$ MeV for
all the 66 Skyrme interactions. For each Skyrme interaction, we
calculate the neutron skin thickness of the nuclei for which the
neutron skin thickness data from antiprotonic atom measurement are
available. Then we get the parameters $ C_{Z}$ and $C_{\delta}$ in
Eq.(3) for each Skyrme interaction. The relationships between the
slope parameter $L$ and $ C_{Z}$, $C_{\delta}$ thus can be obtained.
Fig.4 displays the correlations between the slope parameter $L$ of
the symmetry energy and the parameters $ C_{Z}$ (left panel) and
$C_{\delta}$ (right panel). The full circles are the calculated $
C_{Z}$ and $C_{\delta}$ for each Skyrme interaction. From the left
panel we can see that the parameter $ C_{Z}$ changes from $-0.006$
to $-0.003$ fm which is totally in the range of $[-0.0068,-0.0014]$
fm extracted from the experimental data, and shows an
anti-correlation with $L$ of Skyrme interactions. In the right panel
of Fig.4, a clearly linear increasing correlation between $L$ and
$C_{\delta}$ can be observed. The slope parameters $L$ for different
Skyrme interactions varies with $C_{\delta}$ in an area limited by
two lines $L = (-117.93 \pm 2\sigma)+151.21 C_{\delta} $ ($\sigma =
6.54$) which are plotted in Fig.4(b) with dashed red lines. With the
range of $C_{\delta} = (1.05 \pm 0.08)$fm extracted from the
experimental data, the slope parameter $L$ is constrained in $16
\lesssim L \lesssim 66$ MeV. Within the 66 Skyrme interactions we
adopted, the Skyrme interactions satisfying this relation are SLy
series, SkM, SkM$^{*}$, MSk1$\sim$2, Skz0$\sim$1, SkT3, SkT6$\sim$9,
SII, SIII$^{*}$, SGII, RATP. The corresponding symmetry energy
coefficients at the saturation density $\rho_{0}$ are from $28$ MeV
to $32$ MeV for these selected Skyrme interactions. The uncertainty
of $L$ obtained from the extracted range of $C_{\delta}$ is much
smaller than that from $C_{Z}$. It means that the correlation
between $\Delta R_{np}$ and isospin asymmetry $\delta$ of nuclei is
more sensitive to the density dependence of the symmetry energy than
that between $\Delta R_{np}$ and nuclear charge. The obtained
constraint on the slope parameter $L$ of the symmetry energy is in
consistent with that obtained from the double neutron-proton ratios
and isospin diffusion in heavy ion collisions\cite{Tsang09}.

In summary, we propose an alternative way for constraining the
density dependence of the symmetry energy by means of the relation
of the neutron skin thickness to the nuclear charge and isospin
asymmetry of nuclei. We show that the neutron skin thickness depends
on the isospin asymmetry and the charge of nuclei with the form of
$Z^{2/3}$. By fitting the available data of neutron skin thickness
for a series of nuclei obtained from antiprotonic atom measurement,
the parameters for the correlation between $\Delta R_{np}$ and the
nuclear charge ($C_{z}$) and isospin asymmetry ($C_{\delta}$) can be
extracted. Within the framework of the semi-classical extended
Thomas-Fermi approximation
 together with the Skyrme energy density functional and Coulomb
energy density, we systematically calculate the neutron skin
thickness with 66 Skyrme interactions for the 26 nuclei, for which
the experimental data are available, and get the parameters $C_{z}$
and $C_{\delta}$ for each Skyrme interaction, respectively. Based on
the clearly linear correlation between the slope parameter of the
symmetry energy and the slope parameter $C_{\delta}$ of the neutron
skin thickness, we obtained the constraint on the slope parameter of
the symmetry energy, i.e. $16 \lesssim L \lesssim 66$ MeV, by using
the extracted parameter $C_{\delta}$ from the experimental data. The
Skyrme interactions satisfying this constraint are selected. It
should be stressed that the spherical symmetry Fermi distribution
for proton and neutron density in our calculation is in accordance
with that in the analysis in the antiprotonic measurement. It is
more suitable and consistent to extract the parameters $C_{z}$ and
$C_{\delta}$ with the data from antiprotonic measurement. We suggest
more neutron skin thickness for different element with the same
isospin asymmetry to be measured experimentally, so as to extract a
more accurate dependence of the neutron skin thickness on the
nuclear charge and the isospin asymmetry. With the increasing of the
number and the accuracy of data for neutron skin thickness of nuclei
based on the antiprotonic atom measurement in future, we believe
more accurate constraint on the density dependence of the symmetry
energy can be obtained with our proposed method.

\begin{center}
\textbf{ACKNOWLEDGEMENTS}
\end{center}
    This work is supported by the National Natural Science
    Foundation of China under Grants 10875031, 10847004, 10675172,
    the Doctoral Station Foundation of Ministry of Education of China under Grant 200800270017,
    and the National Basic research program of China under Grants 2007CB209900, 2010CB832903.

\end{document}